# Visualization of Surface-acoustic-wave Potential by Transmission-mode Microwave Impedance Microscopy

Lu Zheng, Di Wu, Xiaoyu Wu, Keji Lai*

Department of Physics, University of Texas at Austin, Austin TX 78712, USA

* E-mail: kejilai@physics.utexas.edu

## Abstract

Elastic waves propagating in piezoelectric materials are accompanied by a time-varying electric potential, which is of critical importance for acousto-electronic applications. The spatial mapping of such a potential at microwave frequencies is challenging since the characteristic length scale is determined by the acoustic wavelength of several micrometers. In this work, we report the visualization of surface acoustic waves (SAWs) on ferroelectric samples by transmission-mode microwave impedance microscopy (T-MIM). The SAW potential launched by the interdigital transducer is detected by the tip and demodulated by the microwave electronics as time-independent spatial patterns. Wave phenomena such as interference and diffraction are imaged and the results are in excellent agreement with the theoretical analysis. Our work opens up a new avenue to study various electromechanical systems in a spatially resolved manner.

The linear electromechanical coupling in piezoelectric materials enables the interconversion between electrical and acoustic signals[1], which has found numerous applications in modern science and technology[2,3]. For instance, the excitation and detection of surface acoustic waves (SAWs) in quartz crystals are widely utilized in electronic components such as delay lines, filters, and oscillators[2,3]. The oscillating potential carried by propagating SAWs can manipulate the two-dimensional electrons hosted in piezoelectric GaAs quantum wells[4,5]. In periodically poled superlattices of piezoelectric/ferroelectric lithium niobate ($LiNbO_3$), the coupling between giga-Hertz (GHz) electromagnetic and acoustic waves leads to polaritons with novel phononic band structures[6-11]. In these systems, the characteristic dimension is set by the acoustic wavelength in piezoelectric solids, which is 5 orders of magnitude smaller than the electromagnetic wavelength at the same frequency ($f$). Research work that provides spatial information in the mesoscopic length scale is therefore highly desirable for studying wave phenomena, such as the interference, diffraction, and localization, of few-GHz supersonic SAWs.

In the past few decades, much effort has been made to probe acousto-electronic properties in a spatially resolved manner. Scanning laser interferometry, for example, images the out-of-plane displacement fields[12,13] with sub-picometer sensitivity and a diffraction-limited lateral resolution around 1 μm. The spatial resolution of surface displacement fields can be improved to ~ 20 nm in scanning acoustic force microscopy (SAFM), which detects the nonlinear mixing of two slightly detuned SAWs through a cantilever probe at the difference frequency[14-16]. The sub-nm SAW amplitude has also been visualized by stroboscopic X-ray imaging[17,18]. On the other hand, the piezoelectric SAW potential that is critical for the applications has not been thoroughly studied. It was demonstrated that the secondary electrons in a scanning electron microscope (SEM) could be modulated by the spatially varying SAW electric field[19,20]. The applicability of this method, however, is rather limited due to the strong charging effect in insulating piezoelectric crystals[19], resulting in a moderate resolution of ~ 1 μm and an operation frequency below 0.5 GHz. In this paper, we report the visualization of piezoelectric SAW potential on the surface of $z$-cut $LiNbO_3$ crystals by transmission-mode microwave impedance microscopy (T-MIM), an atomic-force-microscopy (AFM) based technique with sub-100 nm spatial resolution. The SAW potential generated by interdigital transducers (IDTs) is demodulated by the homodyne detection electronics, showing time-independent spatial patterns in the two orthogonal channels. The superposition of two counter-propagating SAWs and wave diffraction due to a small domain with opposite polarization are also observed by the T-MIM. Our work introduces a new direction to locally probe acousto-electronic phenomena in complex systems by near-field electromagnetic imaging.

As a rapidly evolving technique in recent years, MIM is commonly used to study the nanoscale permittivity and conductivity distribution in advanced materials[21-24]. In a typical reflection-mode MIM (R-MIM) setup shown in Fig. 1a, the GHz signal is delivered to the probe through an impedance-match section[22]. The small variation of tip-sample impedance $Z_{t-s}$ during the measurement leads to changes of the reflected microwave, which is demodulated by the quadrature IQ mixer. By adjusting the local oscillator (LO) phase $\phi$, the real and imaginary components of the admittance change $\Delta Y_{t-s} = \Delta(Z_{t-s}^{-1})$ can be mapped as R-MIM-Re/Im images[22]. As shown in Fig. 1b, it is straightforward to reconfigure the system as a transmission-mode MIM (T-MIM), where the tip acts as a receiver to detect the local RF voltage $V_s$. The equivalent circuits of R-MIM and T-MIM are schematically shown in Figs. 1c and 1d, respectively. At our operation frequency of ~ 1 GHz, the cantilever probe[25] can be viewed as a lumped element dominated by an effective capacitance of 1 pF. Using transmission-line analysis[26] (Appendix A), it can be shown that the receiver has an input impedance $|Z_{in}|$ ~ 1 kΩ at 1 GHz. Through a similar tip-sample coupling impedance $Z'_{t-s}$, an input signal of $V_{in} = V_s \cdot Z_{in}/(Z'_{t-s} + Z_{in})$ is picked up by the tip and then amplified and demodulated by the electronics. It is worth noting that similar transmission-type probes have been used to map out the RF fields in microwave resonators[27,28] and metamaterials[29,30]. In those systems, however, the characteristic length scale is determined by the electromagnetic wavelength (30 cm at 1 GHz) and a mesoscopic spatial resolution is not necessary.

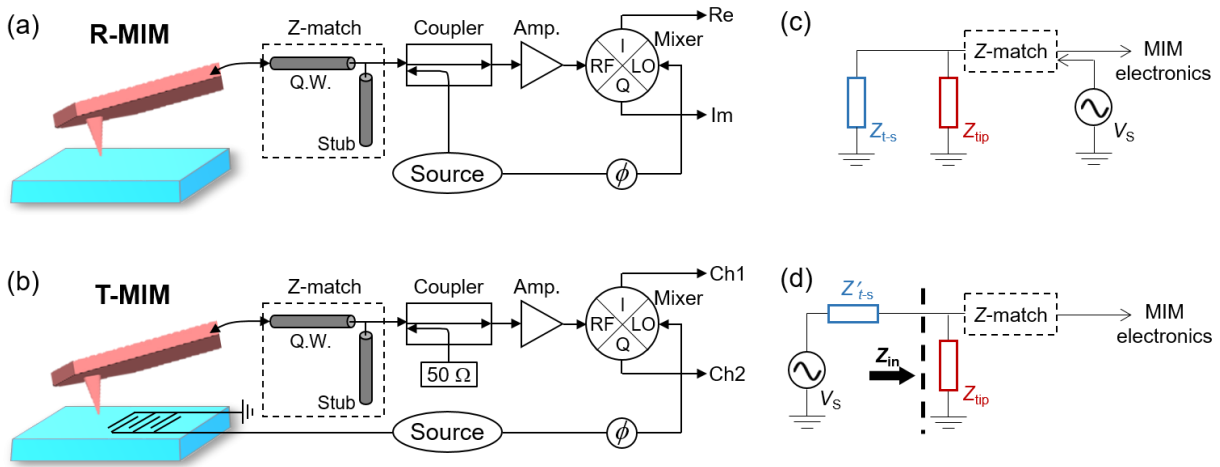

FIG. 1. (a) Schematic of the R-MIM. The excitation signal is delivered to the tip and the reflected signal is amplified and demodulated by the IQ mixer to form the R-MIM-Re/Im images. (b) Schematic of the T-MIM. The excitation signal is delivered to the IDT on the sample and the transmitted signal is amplified and demodulated by the IQ mixer to form the T-MIM-Ch1/Ch2 images. (c) Equivalent circuits of the R-MIM and (d) T-MIM.

Fig. 2a shows the SEM image of a pair of IDTs used in our experiment. The device was designed to excite the *x*-propagating Rayleigh-type SAW on the *z*-cut LiNbO$_3$ substrate, which was poled to be a single ferroelectric domain prior to the device fabrication. LiNbO$_3$ has a trigonal crystal structure with a mirror *yz*-plane and a direct triad *z*-axis along the polar direction[31]. Using finite-element modeling (Appendix B), one can show that the piezoelectric SAW potential $V_s$ is about 10% of the excitation voltage at the transmitting IDT (±1 V). In addition, since the coupling impedance $Z'_{t-s}$ between the tip and metal electrodes is much smaller than that between the tip and LiNbO$_3$, the signals on the IDTs are very strong and saturate the T-MIM output. The S-parameters of the two IDTs measured by a vector network analyzer are plotted in Fig. 2b. The passband of ~ 50 MHz around 1 GHz is consistent with the use of 20 pairs of interdigital fingers[2]. The dip of $S_{12}$ in the middle of the passband is likely due to the SAW reflection from the receiving IDT.

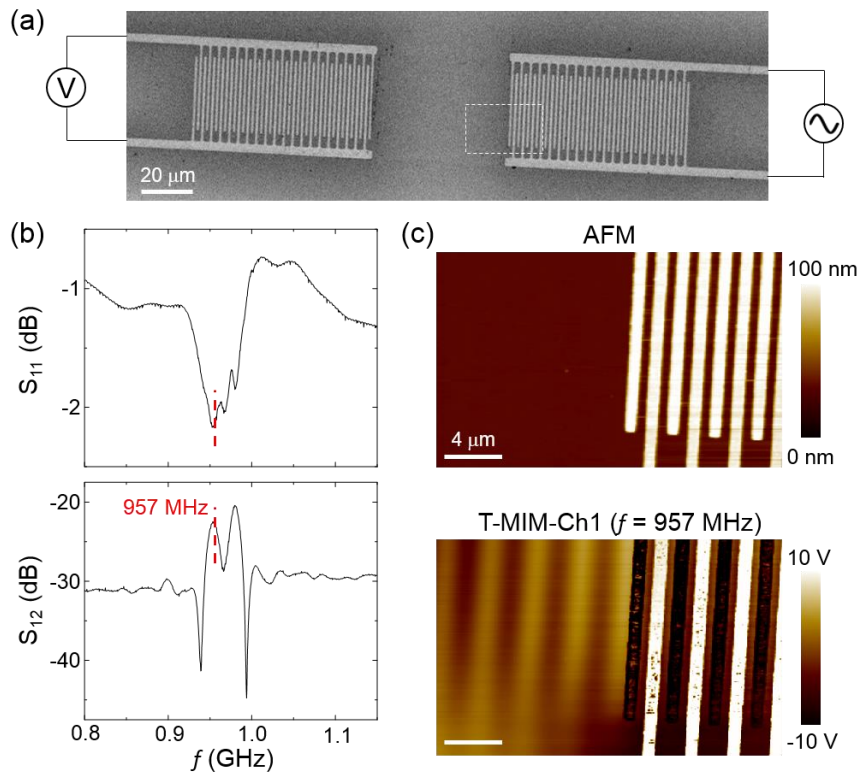

FIG. 2. (a) SEM image of the SAW device with a pair of IDTs. (b) Return loss $S_{11}$ and insertion loss $S_{12}$ of the SAW device measured by a vector network analyzer. The T-MIM frequency of 957 MHz is labeled in the plots. (c) AFM and T-MIM images in the dashed rectangular region in (a). Wave-like features are seen in the T-MIM data. The scale bars are 4 μm.

Fig. 2c displays the simultaneously acquired AFM and T-MIM-Ch1 images when the excitation IDT is powered by 10 dBm microwave at $f = 957$ MHz. While only the interdigital fingers are seen in the surface topography, the electrical potential on both the IDT and the LiNbO$_3$ surface can be clearly imaged by the T-MIM. In Fig. 3, we focus on the data taken in an area of 10 μm × 20 μm between the two IDTs. The featureless R-MIM images in Fig. 3a indicate that there is no permittivity or conductivity variation, whereas the two T-MIM images in Fig. 3b exhibit sinusoidal patterns. As discussed before, the tip is picking up an input signal that is proportional to the SAW potential. Without loss of generality, the signals at the RF and LO ports of the mixer can be represented as $V_{RF} \propto V_s \propto e^{i(\omega t - kx)}$ and $V_{LO} \propto e^{i(\omega t + \phi)}$ (ω: angular frequency, $k$: acoustic wave vector, $\phi$: mixer phase), respectively. Ignoring the terms containing $2\omega t$, we obtain the output signals from the quadrature mixer as follows.

$$V_{Ch1} \propto \mathrm{Re}(V_{RF} V_{LO}^*) = \cos(kx + \phi) \tag{1}$$
$$V_{Ch2} \propto \mathrm{Im}(V_{RF} V_{LO}^*) = -\sin(kx + \phi) \tag{2}$$

In other words, the electronics demodulate the time-varying SAW potential into time-independent spatial patterns, which is in good agreement with the T-MIM data. The line profiles in Fig. 3c show the same amplitude and a phase difference of 90° between the two channels. We have also confirmed that a change in the mixer phase $\phi$ introduces the same phase shift to both channels. By fitting the periodicity of the sinusoidal curves, a phase velocity of $v = 3.8$ km/s is obtained, which is consistent with that of the $x$-propagating Rayleigh SAW[32]. Further analysis of the signal level (Appendix A) also shows that the tip-sample coupling impedance in our experiment $|Z'_{t-s}| \sim 160$ kΩ, which is much greater than $|Z_{in}|$.

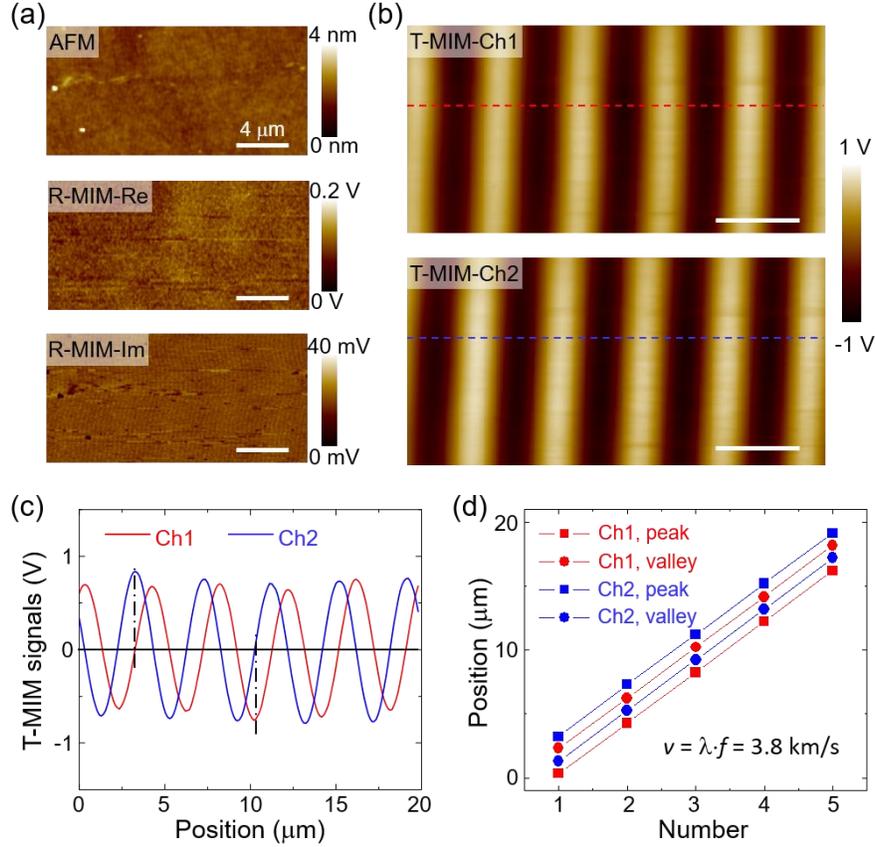

FIG. 3. (a) AFM and R-MIM-Re/Im images in an area between the two IDTs. (b) T-MIM-Ch1/Ch2 images in the same area as (a). All scale bars are 4 μm. (c) Line profiles of the two T-MIM channels. The dash-dotted lines show that the two sinusoidal curves are offset by 90°. (d) Positions of the peaks and valleys in the T-MIM data. The linear fits to the data points correspond to the Rayleigh SAW speed of 3.8 km/s.

We now turn to the T-MIM imaging of a standing wave formed by two counter-propagating SAWs. Fig. 4a shows the schematic of the experimental setup, where two balanced signals (0 dBm in amplitude) with a phase offset of $\theta$ are fed into the pair of IDTs. This geometry is technologically important in that it can create acoustic trapping potentials for electrons[33]. Following the same analysis above, the input signals to the mixer can be written as $V_{RF} \propto e^{i(\omega t - kx)} + e^{i(\omega t + kx + \theta)}$ and $V_{LO} \propto e^{i\omega t}$. The LO phase $\phi$ is omitted since it contributes the same phase to both channels. The mixer then generates two output signals as follows.

$$V_{Ch1} \propto \text{Re}(V_{RF}V_{LO}^*) = \cos kx + \cos(kx + \theta) \tag{3}$$
$$V_{Ch2} \propto \text{Im}(V_{RF}V_{LO}^*) = -\sin kx + \sin(kx + \theta) \tag{4}$$

By tuning the phase difference $\theta$ between the two counter-propagating SAWs, the signal levels of the two T-MIM channels can be varied. When $\theta = 0°$, the sinusoidal spatial patterns are expected to appear only in Ch1 ($V_{Ch1} \propto 2\cos kx, V_{Ch2} \propto 0$). The patterns should then be the same in both channels when $\theta = 90°$ ($V_{Ch1} = V_{Ch2} \propto \cos kx - \sin kx$) and completely move to Ch2 when $\theta = 180°$ ($V_{Ch1} \propto 0, V_{Ch2} \propto -2\sin kx$). As seen in Fig. 4b, the predicted evolution is again in excellent agreement with the measured T-MIM data. The results demonstrate that T-MIM can probe the acoustic standing waves in piezoelectric materials.

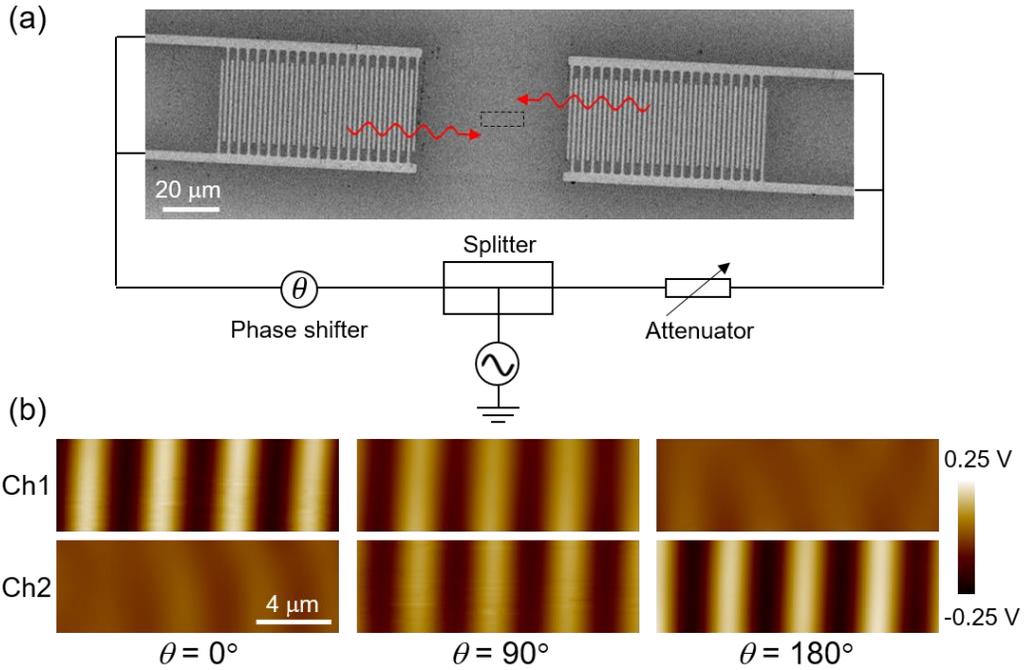

FIG. 4. (a) Schematic diagram for the imaging of counter-propagating waves. Two signals (0 dBm in amplitude and phase offset by θ) split from the same source are fed into the two IDTs. (b) T-MIM images at different θ's, showing the transition of signal strength from Ch1 to Ch2.

Finally, we briefly discuss the visualization of SAW diffraction due to the presence of a small domain. LiNbO$_3$ wafers poled to be a single ferroelectric domain are energetically unstable. Over an extended period, small domains with opposite polarization may spontaneously form to reduce the electrostatic energy. The domain inversion flips the sign of odd-rank tensors (polarization, 1st rank; piezoelectric tensor, 3rd rank), while leaving the even-rank tensors (permittivity, 2nd rank; elasticity tensor, 4th rank) unchanged[31]. As the acoustic impedance is mostly dependent on the density and elasticity of the material, the SAW displacement field is not strongly affected by the domain structure. In contrast, the SAW electric field, which is the gradient of the SAW potential, changes sign across a domain wall due to the piezoelectric coupling. In our sample, a small domain with spontaneous polarization reversal is found near the left IDT (AFM image in Fig. 5a), as seen in the piezo-force microscopy (PFM) image in Fig. 5b. Since its dimension is comparable to the acoustic wavelength, wave diffraction is expected around the domain. In Fig. 5c, the T-MIM surface potential maps indeed display very strong distortion of the wave front in this region. The line profiles in Fig. 5d further verify that the T-MIM signals from both channels switch sign when passing through the small domain. The capability to resolve the spatial distribution of SAW potential is of particular interest to phononic meta-materials based on periodic ferroelectric domain structures[6-11].

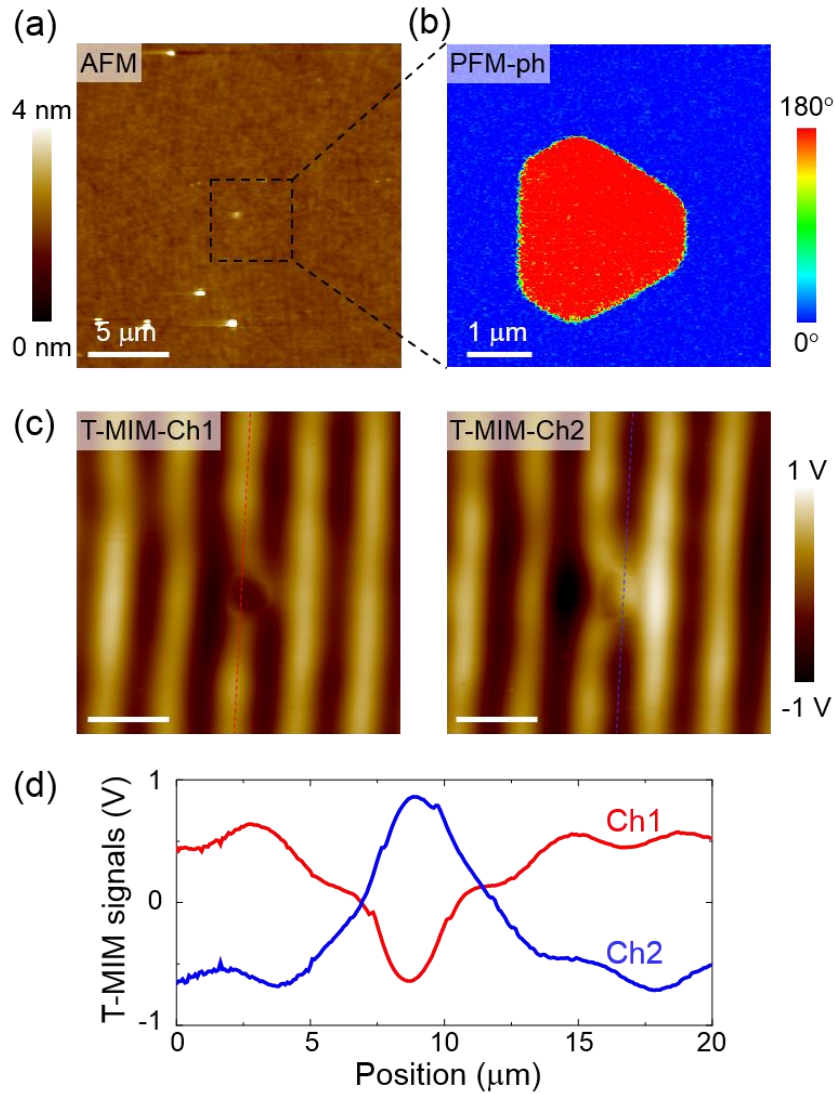

FIG. 5. (a) AFM image in an area with a spontaneously reversed domain. (b) Close-up view of the PFM phase image inside the dashed square of (a). The polarization reversal of the internal domain is evident from the 180°-phase contrast. (c) T-MIM-Ch1/Ch2 images in the same area as (a). Strong distortion of the wave front is seen around the small domain. The scale bars are 5 μm. (d) T-MIM line profiles in (c), showing the sign change of piezoelectric potential in the opposite domain.

To summarize, we demonstrate the visualization of piezoelectric SAW potential on $z$-cut LiNbO$_3$ surface by transmission-mode microwave imaging. The traveling or standing SAW potential generated by IDTs is demodulated by the microwave electronics and mapped as stationary spatial patterns. The signals can be explained by the standard microwave analysis. The wave diffraction due to a spontaneously reversed domain is also seen in the T-MIM images. Our work paves the way to probe nanoscale acousto-electronic behaviors in SAW devices, quantum materials, and phononic crystals.

## ACKNOWLEDGMENTS

This research was supported by NSF Division of Materials Research Award DMR-1707372. The authors thank H. Dong, D. Sounas, and A. Alu for helpful discussions.

## APPENDIX A: Microwave Circuit Analysis

The quantitative analysis of the T-MIM circuit is provided here. The cantilever probe[25] can be modeled as a lumped RLC element with $R_{tip} = 4$ Ω, $L_{tip} = 2$ nH, and $C_{tip} = 1$ pF. At $f = 1$ GHz, the effective tip impedance $|Z_{tip}| \sim 150$ Ω is dominated by the capacitive reactance. As shown in Fig. A1a, an impedance-match network consisting of a quarter-wave cable (AstroLab, Astro-Boa-Flex III, ~ 5 cm) and a tuning stub (Micro-Coax, UT-085C-TP, ~ 5 cm) is needed to route the tip impedance to the 50Ω transmission line[22]. Fig. A1b shows the calculated return loss Γ of the $Z$-match circuit, which agrees with the result measured by a vector network analyzer. Using the standard transmission-line analysis[26], one can then compute the effective impedance viewed from the tip side, i.e., the input impedance $Z_{in}$ of the receiver. For the T-MIM experiment, a large $|Z_{in}|$ is desirable for signal pick-up. As shown in Fig. A1c, $|Z_{in}|$ reaches a maximum of ~ 1 kΩ at the matching frequency, which is used as the operation frequency for both R-MIM and T-MIM.

The tip-sample coupling impedance $Z'_{t-s}$ is estimated as follows. First, the relation between $V_1$ and $V_2$ (Fig. A1a) can be analyzed by considering the forward and backward propagating waves in the quarter-wave cable. On the other hand, these two voltages provide the link between the source signal $V_s$ and the MIM output signal $V_{MIM-out}$.

$$V_1 = V_s \cdot Z_{in}/(Z_{in} + Z'_{t-s}) \approx V_s \cdot Z_{in}/Z'_{t-s} \tag{A1}$$

$$V_2 = V_{MIM-out}/G_{MIM} \tag{A2}$$

Here $G_{MIM}$ = 86 dB is calibrated for the T-MIM electronics. Based on our experimental data, a peak-to-peak SAW potential of 0.2 V (Appendix B) corresponds to a peak-to-peak T-MIM output signal of ~ 2 V (Fig. 3c). The computed tip-sample coupling impedance is plotted in Fig. A1d. The result shows that in this particular experiment, $|Z'_{t-s}|$ is around 160 k$\Omega$, or an effective capacitance of 1 fF, at 1 GHz. Note that $Z'_{t-s}$ strongly depends on the tip apex condition and the sample properties. A small $Z'_{t-s}$ is desirable for efficient T-MIM detection, which, however, usually comes at a price of blunt tip and reduced spatial resolution.

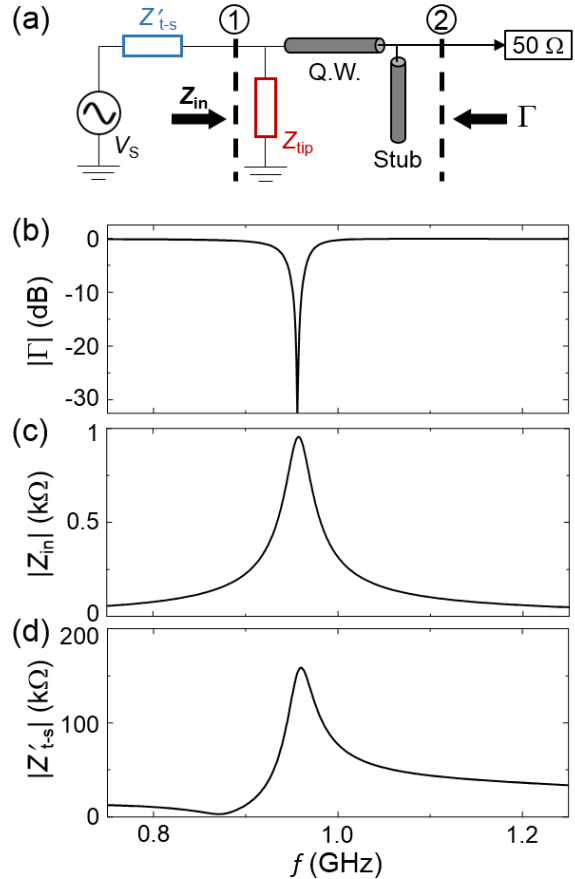

FIG. A1. (a) Equivalent circuit of the T-MIM tip and the Z-match network. (b – d) Simulated return loss of the Z-match circuit, input impedance of the tip as a receiver, and tip-sample coupling impedance. The matching frequency with minimal |Γ| is used as the operation frequency for both R-MIM and T-MIM modes.

**APPENDIX B: Finite-element Modeling**

The piezoelectric SAW potential can be numerically computed by the Structural Mechanics Module in commercial finite-element analysis (FEA) software COMSOL 4.4. Here we simulate a thin plate (1 μm in thickness) with periodic boundary condition along the $y$-direction. The LiNbO$_3$ region, whose permittivity, piezoelectric coefficient, and elasticity tensor are taken from Ref. 31, is bounded by the perfectly matched layer (PML) to avoid wave reflection from the boundary. The IDT spacing is set to be 3.8 μm. Alternating voltage (1 V in amplitude and 1 GHz in frequency) and ground (0 V) are applied on the IDT fingers to excite the $x$-propagating Rayleigh-type SAW on the $z$-cut LiNbO$_3$ surface. Fig. B1a shows the simulated piezoelectric potential distribution in the sample, where the SAW is clearly seen. The surface potential from the simulation (Fig. B1b) indicates that the peak-to-peak SAW potential is 0.2 V, which is the source signal for the T-MIM measurement. Using this information in Eq. (A1), we are able to evaluate the tip-sample coupling impedance $Z'_{t-s}$, which is crucial to understand the signal level in our T-MIM experiment in a quantitative manner.

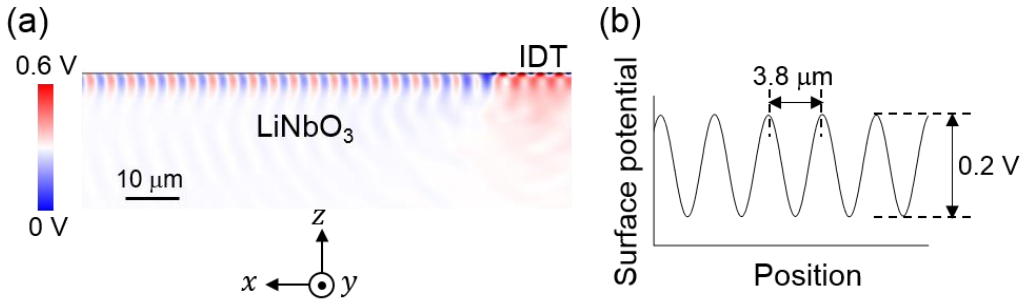

FIG. B1. (a) Piezoelectric potential distribution simulated by finite-element modeling. The voltage on the IDT fingers in this snapshot is 1V/0V. (b) Simulated SAW potential as a function of the position.

**References:**

(1) D. Royer and E. Dieulesaint, *Elastic Waves in Solids*, Springer-Verlag 1999.